\documentclass[
 reprint,
superscriptaddress,
 amsmath,amssymb,
 aps,
prx,
floatfix,
 lengthcheck
]{revtex4-1}
\usepackage[utf8]{inputenc}
\usepackage[T1]{fontenc}
\usepackage[version=4]{mhchem}
\usepackage{graphicx}
\usepackage{placeins}
\usepackage[usenames, dvipsnames]{xcolor}
\usepackage{url}
\date{\today}

\begin{document}
\author{E.\,A.~Eronen}
\email{eemeli.a.eronen@utu.fi}
\affiliation{University of Turku, Department of Physics and Astronomy, FI-20014 Turun yliopisto, Finland}
\author{A.~Vladyka}
\affiliation{University of Turku, Department of Physics and Astronomy, FI-20014 Turun yliopisto, Finland}
\author{F.~Gerbon} 
\affiliation{ESRF, The European Synchrotron, 71 Avenue des Martyrs, CS40220, 38043 Grenoble Cedex 9, France}
\author{Ch.\,J.~Sahle}
\affiliation{ESRF, The European Synchrotron, 71 Avenue des Martyrs, CS40220, 38043 Grenoble Cedex 9, France}
\author{J.~Niskanen}
\email{johannes.niskanen@utu.fi}
\affiliation{University of Turku, Department of Physics and Astronomy, FI-20014 Turun yliopisto, Finland}
\title{Information Bottleneck in Peptide Conformation Determination by X-ray Absorption Spectroscopy
}
\begin{abstract}
	We apply a recently developed technique utilizing machine learning for statistical analysis of computational nitrogen K-edge spectra of aqueous triglycine. This method, the emulator-based component analysis, identifies spectrally relevant structural degrees of freedom from a data set filtering irrelevant ones out. Thus tremendous reduction in the dimensionality of the ill-posed nonlinear inverse problem of spectrum interpretation is achieved. Structural and spectral variation across the sampled phase space is notable. Using these data, we train a neural network to predict the intensities of spectral regions of interest from the structure. These regions are defined by the temperature-difference profile of the simulated spectra, and the analysis yields a structural interpretation for their behavior. Even though the utilized local many-body tensor representation implicitly encodes the secondary structure of the peptide, our approach proves that this information is irrecoverable from the spectra. A hard X-ray Raman scattering experiment confirms the overall sensibility of the simulated spectra, but the predicted temperature-dependent effects therein remain beyond the achieved statistical confidence level.
	
\end{abstract}
\maketitle

\section{Introduction}

Proteins are formed of amino acids, often cited as the building blocks of life. The backbone of the amino acid chain consists of a -CO-NH-C$_\alpha$- sequence, where adjacent residues are bound by the peptide bond \cite{Liljas2009}. The geometry of the backbone contains pairs of flexible dihedral angles, the Ramachandran angles, which together define the macroscopic structure and the function of the protein. The related folding of proteins is a complex and complicated question \cite{Dill2012}, affected by both intramolecular and intermolecular interactions. The process is ultimately dependent on the primary order of the amino acids \cite{Liljas2009} and the surrounding solvent \cite{Ball2017,Canchi2013}.

While diffraction experiments allow for structure determination of proteins, they require a crystalline sample. Owing to its localized mechanism, X-ray spectroscopy maintains its sensitivity to atomistic structure also in the soft condensed phase, although statistical variation of the individual spectra in such environments has been found to be huge \cite{Wernet2004, Ottosson2011, Niskanen2016a, Niskanen2016b, Niskanen2017, Sahle2018, Niskanen2019, VazDaCruz2019, Pietzsch2022}. This variation can be accounted for by evaluation of the spectra at numerous structures, sampled from the respective statistical ensemble, and averaging over them. Furthermore, X-ray spectroscopy is sensitive to the local structure of amino acids and peptides \cite{Kaznacheyev2002,Gordon2003,Zubavichus2004,Zubavichus2005,Messer2005,Ottosson2011,Blum2012,Weinhardt2019}. For example, different protonation forms of glycine yield different resonant inelastic X-ray scattering spectra \cite{Blum2012} and X-ray photoelectron spectra \cite{Ottosson2011,Niskanen2013}. It has also been found that near-edge X-ray absorption fine structure is sensitive to a few neighbouring amino acids in poly-Gly peptides \cite{Gordon2003,Zubavichus2004,Weinhardt2019}.

To predict the native form of proteins in their biological environment, classical molecular dynamics (MD) is often used \cite{Hollingsworth2018}. In this technique, the atomistic composition of the system is accounted for, but the forces between the atoms are computed from a prior parametrization rather than from the electronic-nuclear system that ultimately defines them. Being computationally much lighter, and allowing for time and size scales not otherwise accessible, this makes classical MD subject to the quality of the parametrization -- the force field -- used in the calculations \cite{LindorffLarsen2012,MartinGarcia2015}. For the development of these models in the solution environment, their performance assessment in the atomistic level would be valuable.

Interpretation of X-ray spectra is typically not straightforward. This is due to the complicated relation between spectra and structures, originating from quantum nature of the electronic system. In addition, the statistical aspects of the liquid state cause an experiment to probe only the ensemble average. To tackle these problems, machine learning and related emulator-based component analysis (ECA) \cite{Niskanen2022} may prove useful. The ECA algorithm carries out dimensionality reduction in structural space to identify structural variations having the strongest influence on the spectra. Thus the method identifies recoverable (and irrecoverable) structural information without a prior hypothesis.

\begin{figure}
	\centering
	\includegraphics[width=\columnwidth]{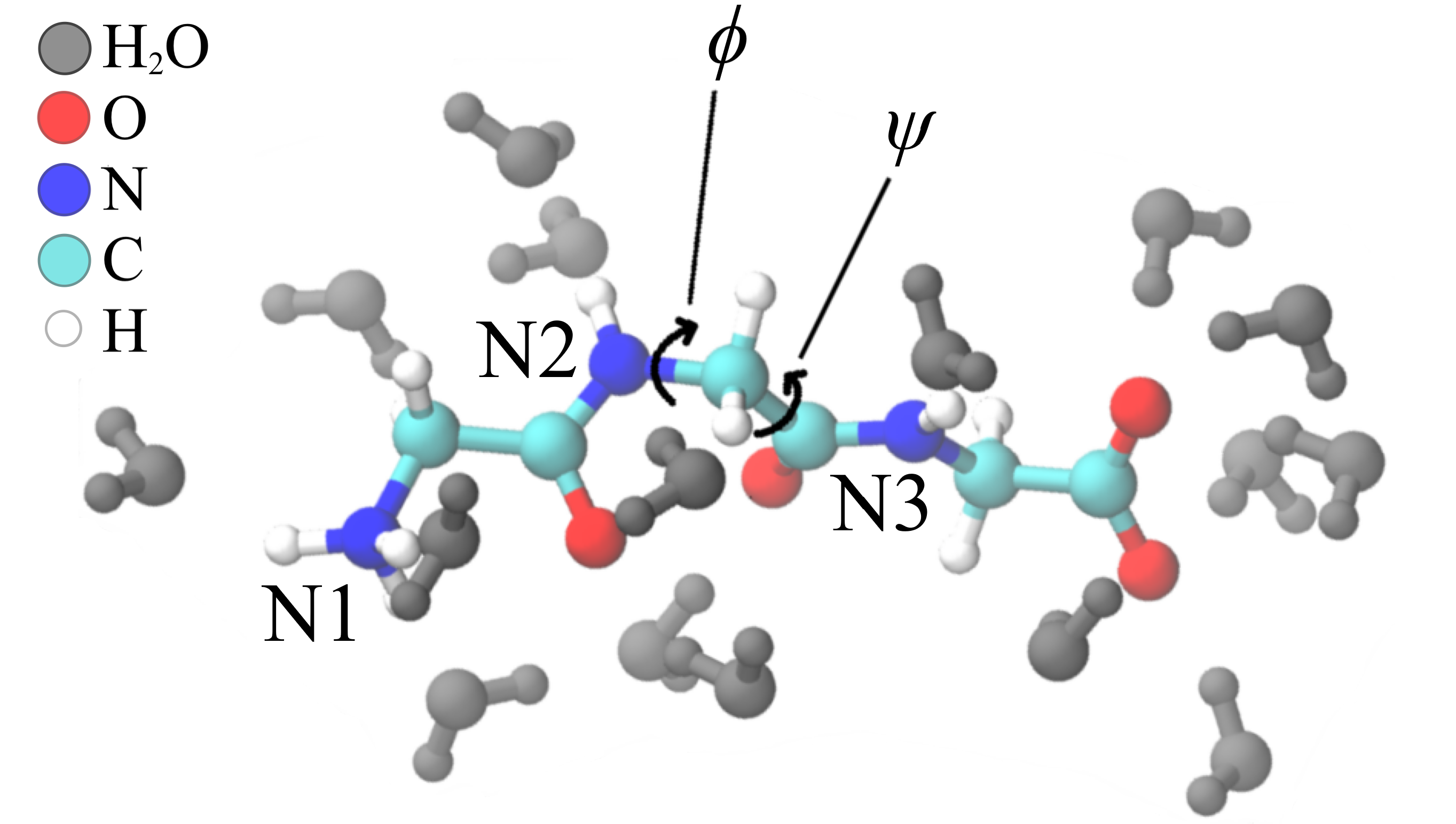}
	\caption{A schematic illustration of aqueous triglycine. The absorption sites are denoted as N1, N2 and N3. The backbone dihedrals known as the Ramachandran angles $\phi$ and $\psi$ are shown for the central residue. The molecular plot was prepared using the VMD software \cite{HUMP96}.}
	\label{fig:trigly}
\end{figure}

In this work we report a theoretical study of the smallest tripeptide triglycine in aqueous solution (see Figure~\ref{fig:trigly}) at temperatures of 300\,K and 350\,K, expecting a change in the distribution of the Ramachandran angles. We calculate an extensive set of N K-edge X-ray absorption spectra from MD trajectories with three different force fields, observing notable structural and spectral variation. Next, we study the dependency between intensities of spectral regions of interest (ROI) and the corresponding structures using a neural network (NN) and ECA. The results show that while X-ray absorption ROIs are sensitive to the nearest neighboring atoms of the absorption site, they are not sensitive to the secondary structure {\it i.e.} the Ramachandran angles of the system, in agreement with Schwartz {\it et al.} \cite{Schwartz2010}. The simulated spectra are in good agreement with an X-ray Raman scattering (XRS) experiment we performed. However, the predicted spectral difference profile as a function of temperature remains unconfirmed with the achieved statistical uncertainty.

\section{Methods}
\subsection{Simulations}
We simulated the NPT ensemble by classical MD at 300\,K and at 350\,K and 1 bar using three different force fields: AMBER-03 \cite{Duan2003}, Charmm27 \cite{MacKerell1998,MacKerell2004} and 
OPLS-AA \cite{Kaminski2001}. 
For the first two force fields we applied the TIP3P \cite{Jorgenssen1983} water model and for the latter the TIP4P \cite{Jorgenssen1983} model. The cubic simulation cell had an edge length of $\sim 50$\,{\AA}. The input files were prepared using a desktop installation of GROMACS \cite{Abraham2015} package (version 2020.1), and the simulations were performed using the version 2020.5 \cite{gromacs2020.5} on a computing cluster. Rigid water molecules were used whereas no constraints were applied on the triglycine molecule. After 1\,ns of initial thermalization with Berendsen thermostat and barostat \cite{Berendsen1984}, we ran MD for 51\,ns (timestep 0.5\,fs) using the Nos\'e-Hoover thermostatting \cite{Nose1984,Hoover1985} (time constant $\tau_T$=0.5\,ps) and Isotropic Parrinello-Rahman barostatting \cite{Andersen1980,Parrinello1980} (time constant $\tau_P$=5.0\,ps,  compressibility $4.5\times10^{-5}$ bar$^{-1}$). To avoid the ``hot-solvent/cold-solute'' problem, we used separate thermostats for the solute and for the solvent in the latter run, with conserved energy drift of the order 1\,kJ\,mol$^{-1}$ns$^{-1}$atom$^{-1}$. The last 50\,ns of the run were sampled with 10\,ps spacing for spectrum calculations. 

From the MD snapshots we calculated the N K-edge X-ray absorption spectra of 30006 structures using the projector-augmented-wave (PAW) method \cite{Enkovaara2010} with plane wave basis and density functional theory (DFT), as implemented in GPAW version 22.1.0 \cite{Mortensen2005,Enkovaara2010,Larsen2017}. As the simulation cell of MD is too large to be used in the quantum mechanical electron structure calculations, we used cut structures including only the water molecules within 3.0\,{\AA} of the solute. A vacuum with a radius of 3.0\,{\AA} was added around each structure. Excitations to the lowest 1500 valence single-electron states were evaluated in the transition potential half-hole (TP-HH) approximation \cite{Triguero1998} for each nitrogen site (N$_\mathrm{ex}$). The spectrum onset was corrected using $\Delta$-DFT method for the lowest core-excited state. The calculations utilized plane wave basis with the energy the cutoff of 350\,eV and the Perdew--Burke--Ernzerhof (PBE) functional \cite{pbe}. To aid convergence, occupation smearing by the Fermi-Dirac distribution (width 0.25\,eV) was used.

The simulations resulted in energy--intensity pairs for the transitions of each absorption site (denoted as N1, N2 and N3; see Figure~\ref{fig:trigly}), from which the spectra were obtained by convolution with Gaussian functions of increasing width analogously to the procedure presented in \cite{Leetmaa2010}. The full width at half maximum of the function was obtained by a numerical grid search for the best match with the experiment; we used 1.4\,eV for the lowest state and increased the value linearly in energy to 4.5\,eV for states 5.5\,eV above it, or higher. An alternative set of convolution parameters was tested (from 0.2\,eV to 4.25\,eV for states 10\,eV above the lowest state, or higher) without qualitative changes in the results (see Supplementary Information). The spectrum of a single snapshot was evaluated as the sum of the spectra from the three nitrogen atoms and the resulting ensemble mean spectra of all the systems matched the experiment well as seen in Figure~\ref{fig:ramasimexp}. In this set, 11 spectra were omitted in the subsequent analysis due to obviously nonsensical results.

To validate the computational results, sets of a few hundred to a few thousand spectra were calculated while varying one of the following simulation parameters: the number of water molecules included (both with a 6.0\,{\AA} cutoff, and also without any water), the plane wave energy cutoff (600\,eV) or the functional (RPBE \cite{rpbe}). Some of these calculations were troubled with convergence issues, but the successful ones show none of the parameter changes to have a significant effect on the temperature difference profile (see Supplementary Information). In total, the spectrum calculations took approximately 300 000 CPU hours on Intel Xeon Gold 6148 processors.

\subsection{Data analysis}

We divided the spectra into three ROIs, the selection of which was based on mean zero-passing location in the T-difference profiles $\Delta$ of the three models. These areas roughly correspond to pre peak (I), main edge (II), and continuum (III) in the spectrum, and each of them has a consistent T-dependence in the simulations. Using the full spectrum instead of ROIs would probably allow more information to be captured, but the risk of overinterpretation also grows with the grid tightness, as simulations are always erroneous in reproducing the experiment \cite{Niskanen2022a}. To conclude, we analyze the dependency between the ROI intensities $\mathbf{S}$ and the corresponding structure $\mathbf{R}$.

We apply emulator-based component analysis \cite{Niskanen2022} to this data for its interpretation. The algorithm searches for a few orthogonal basis vectors of structural space by maximizing the generalized covered variance (R$^2$-score) of the prediction for projections of the data onto the spanned subspace. This procedure utilizes the known target values of the original data points in the evaluation of the score. The components of the ECA basis vectors indicate dominant structural features in terms of the variation of the resulting output (in this work the spectral ROIs). The method needs a suitable machine learning (ML) based emulator $\mathbf{S}_\mathrm{emu}$ to predict the intensities of the three ROIs of any given structure $\mathbf{R}$. We used an emulator consisting of nine separate NNs, one for each nitrogen atom and ROI, implemented by scikit-learn \cite{scikit-learn}.

The individual structures $\mathbf{R}$ were encoded using the local many-body tensor representation (LMBTR) \cite{Huo2022} implemented in the DScribe package \cite{Himanen2020}. The system is represented as distances between pairs and angles between triples of each element combination that includes the central atom -- the absorption site. The internal hyperparameters of the LMBTR descriptor and the emulator were chosen by an alternating search, for best average ROI prediction (see Supplementary Information). In the search for the best descriptor, some LMBTR features were zero depending on the according set of hyperparameters. These features were manually removed before training the final model, without a significant effect on performance. The total dimensionality of the final LMBTR vector $\mathbf{D}(\mathbf{R})$ describing the local neighborhoods of the three nitrogen atoms was $1140$. The ECA algorithm aims at dimensionality reduction, an optimization task complicated by the notably large number of dimensions of this space. We applied an alternating partial optimization with respect to 10\% of the ECA vector components at a time. The routine was completed with a full optimization of all components at once. For these tasks we used the optimization toolbox of the SciPy \cite{Scipy} Python library and the trust-region interior point method \cite{Byrd1999} therein.

For the analysis of general spectrum--structure relationships, we combined all the data from the three force fields and the two temperatures into one data set, which was randomly divided for model selection and training (80\%) and for testing and application (20\%) of the emulator.

\subsection{Experiment}
We measured XRS spectra \cite{schulke2007book} at the N K-edge of aqueous triglycine using the multi-element XRS endstation \cite{Huotari2017} at beamline ID20 at the European Synchrotron Radiation Facility (ESRF). In the experiment scattering signal proportional to the double differential scattering cross section (DDSCS) \cite{Sahle2014a} $\mathrm{d}^2\sigma/\mathrm{d}\Omega\mathrm{d}\omega_2$ at finite angle element $\mathrm{d}\Omega$ and outgoing photon energy element $\mathrm{d}\omega_2$ is recorded with a monochromatic and collimated incident beam of hard X-rays. When carried out at small scattering angles (forward direction) the XRS core-level signal is reduced to the dipole spectrum, and thus yields a spectrum equivalent to XAS \cite{theory1967}. Four analyzer modules, 12 spherically bent Si(660) crystals (bending radius $R_b =1$\,m) in each, were used to detect scattering in near-forward direction. We used the spacial imaging property of the bent analyzer crystals to include signal only from the interaction region by selecting the corresponding pixels of the CCD detector in the analysis. We observed different signal from the bulk liquid region and from the capillary walls and therefore excluded the latter pixels from the analysis.

To minimize the effect from radiation damage, a heatable liquid flow cell (modified version of a cell described elsewhere \cite{sahle2015miniature}) was used. In the system, a magnetically driven pump rotor is used to circulate $\sim$\,5\,ml of sample liquid through a capillary (2\,mm outer diameter, 0.01\,mm wall thickness), where inelastic X-ray scattering takes place. Moreover, the sample was replaced regularly to further reduce possible degradation. The sample solution of molality 0.3\,mol/kg of aqueous triglycine was prepared by dissolving the powder sample (Sigma--Aldrich, purity $\geq$ 99.0\%, lot \#\,BCCB1590) into de-ionized water ($\rho\approx 18.2\,\mathrm{M}\Omega\,\mathrm{cm}$, $\mathrm{TOC}\approx2$ ppb).

\section{Results}

\begin{figure}
	\centering
	\includegraphics[width=\columnwidth]{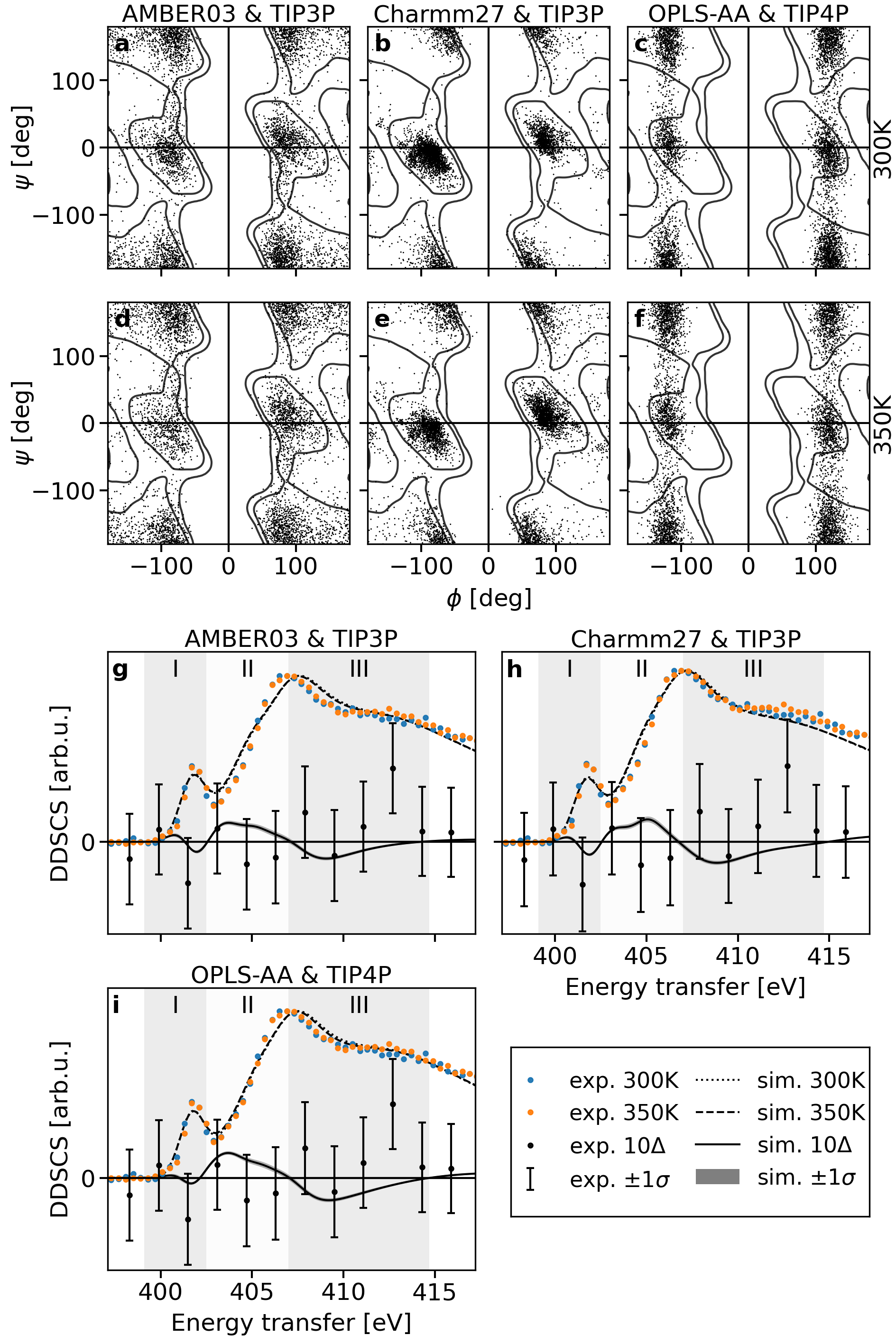}
	\caption{The calculated and experimental results for aqueous triglycine. {\bf a}--{\bf f}: Scatter plot of the Ramachandran angles of the central residues from the simulated trajectories and the definitions \cite{Lovell2003} of the allowed (99.95\%) and favored (99.8\%) regions of glycine residue according to the Top8000 data set \cite{Hintze2016}. {\bf g}--{\bf i}: Computational ensemble mean and experimental spectra (background removed) with the respective temperature difference profiles $\Delta$ and spectral regions of interest. For the experiment, the difference profile has been 4-fold binned from the spectra. The error bars and shading of the simulated curve indicate the statistical uncertainty $\sigma$ (confidence level 68\%). The computational spectra have been shifted by $-$2.9\,eV in all cases for the pre peak to match with the experiment. The experiment is presented scaled for the same main peak height as the respective 300\,K simulation.}
	\label{fig:ramasimexp}
\end{figure}
The Ramachandran angles of the central residue,  $\phi$ and $\psi$ (see Figure~\ref{fig:trigly}), can be visualized in a corresponding scatter plot. Depending on these two dihedrals, the residue can either exhibit an $\alpha$-helix or $\beta$-sheet secondary structure. Our MD simulations show triglycine appears in both of these structural classes (Figure~\ref{fig:ramasimexp}a-f), but the distributions are different between the different force fields. As the molecule has no restrictive side chains, the plot is symmetric for the left-handed and right-handed isomers. The conformational variation of the liquid system gives rise to significant spectral variation. Using simulations allows studying the dependency between the shape of a spectrum and the underlying structure.

Figures \ref{fig:ramasimexp}g-i present the experimental spectra as well as the simulated ensemble-mean spectra for the corresponding force fields and the two temperatures. The temperature difference profile $\Delta$ is also shown. The simulated curves were shifted by $-2.9$\,eV to match the pre peak of the experiment, and the experiment was scaled for the same main peak height with the respective 300\,K simulation. Due to the huge Compton scattering background, approximately subtracted from the spectra, our experiment suffers from a rather large statistical uncertainty. This prohibits further conclusions as the predicted difference profiles lie within. However, an impressive match between the simulated and experimental spectra is obtained.

\begin{figure}
	\centering
	\includegraphics[width=\columnwidth]{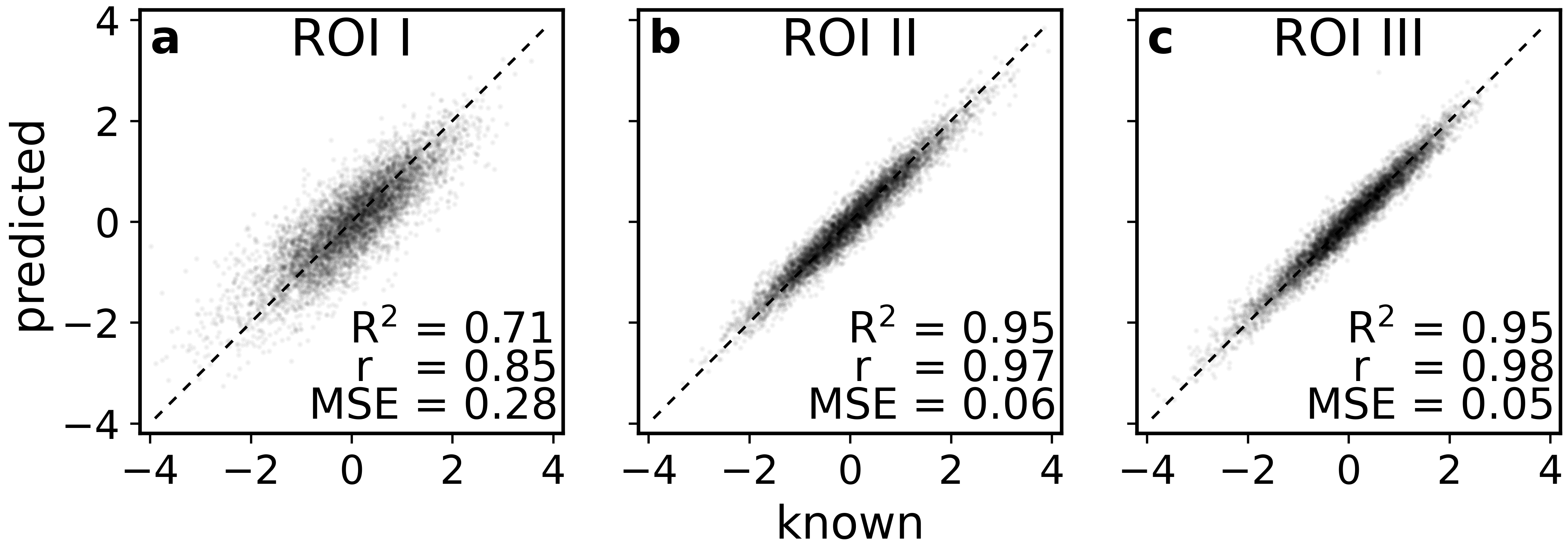}
	\caption{Emulator evaluation with standardized data. {\bf a}-{\bf c}: Predicted intensities for ROIs I--III, respectively, plotted against the known ones. Covered variance R$^2$, Pearson's r, and the mean squared error MSE are given for each panel. Overall covered variance of the three ROIs is 0.874.}
	\label{MLP}
\end{figure}

We did not observe a significant linear correlation between any of the structural features (internal coordinates and water related parameters) and the ROI intensities (see Supplementary Information). However, a well-functioning emulator will enable the use of ECA decomposition as a more advanced analysis tool. In this case, the apparent collaborative action of the structural degrees of freedom is, indeed, captured by an NN-based emulator as depicted in Figure~\ref{MLP}. The emulator achieved covered variances (R$^2$ score) of 0.714, 0.945, and 0.953 for each region I--III with the test data. The prediction accuracy is systematically worse for region I, the origin of which is the N1s$\rightarrow\pi^*$ resonance of the N atoms of the peptide bonds. Combining the regions together, the total covered variance stood at 0.874 for z-score standardized ROIs and 0.942 for the absolute intensity ROIs. The difference can be attributed to much larger total areas of ROIs II and III dominating the absolute score. Going forward, a score of 0.874 is the upper limit for covered spectral variance for the standardized ROI intensities by ECA decomposition, as even with complete structural coverage the respective ML-emulator-induced error will remain.

We split the test data further into two parts of the same size for the ECA. The first part was used for the optimization of the standardized structural descriptor space basis vectors (fit) and the latter was used for an independent test (validation). The covered variance as a function of the rank of the ECA expansion are given in Table \ref{tab:ECAvariances} for both data parts. While the covered ROI intensity variance improves up to rank five (emulator limit is achieved), convergence for the validation data is practically reached with three components.

\begin{table}[h]
	\centering
	\caption{Covered ROI intensity variances as a function of the rank of the ECA decomposition.}\label{tab:ECAvariances}
	\begin{tabular}{ccc}
		Rank & ECA & ECA \\
		& fit & validation \\
		\hline
		1   & 0.536 & 0.519 \\
		2   & 0.753 & 0.712  \\ 
		3   & 0.858 & 0.813 \\
		4   & 0.865 & 0.817 \\
		5   & 0.877 & 0.817 \\
		\hline
		\hline
	\end{tabular}
\end{table}

\begin{figure*}[]
	\centering
	\includegraphics[width=\textwidth]{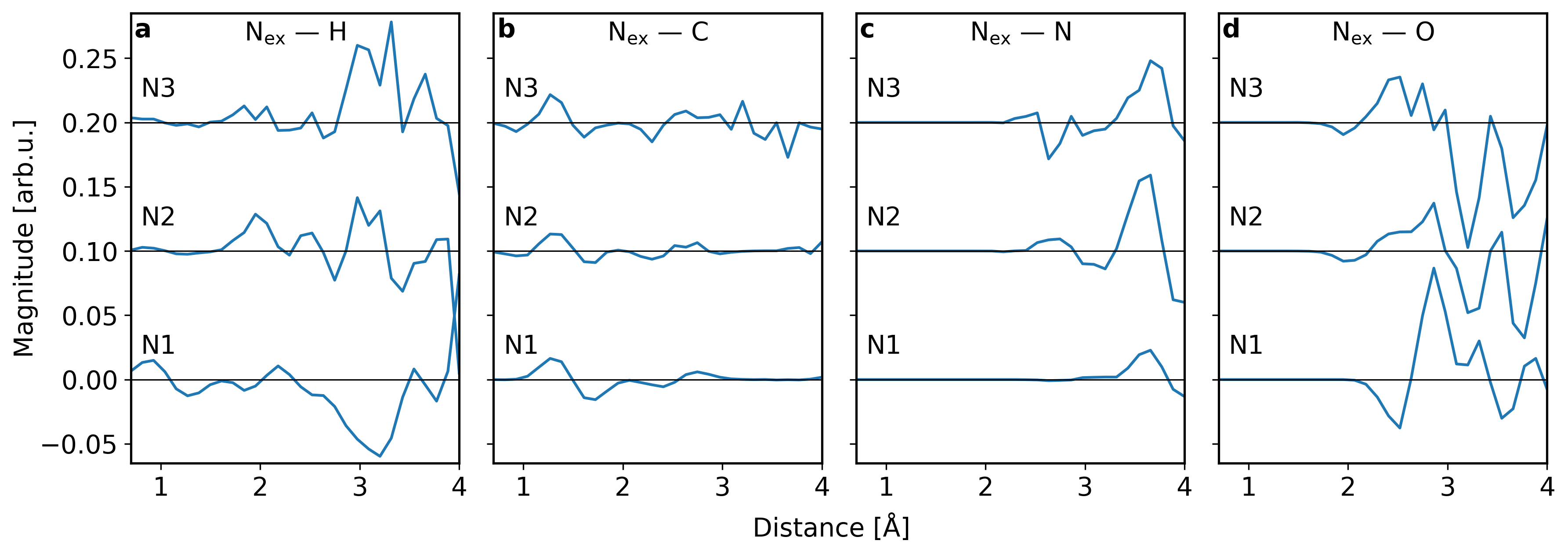}
	\caption{The interatomic distance distributions for each of the nitrogen atoms deduced from the first component vector transformed into the descriptor space. {\bf a}--{\bf d}: Distances from the absorption site (denoted as N$_\mathrm{ex}$) to the neighboring hydrogen, carbon, nitrogen and oxygen atoms, respectively. The ROIs of the total N K-edge spectrum is sensitive to the hydrogen and oxygen atoms at around 3\,{\AA}, which can mainly be contributed to the water molecules. In addition, the ROIs are sensitive to the nearest carbon and nitrogen atoms of each absorption site. Finally, the ROIs are also sensitive to the three nearest hydrogen atoms of N1.
	}
	\label{fig:ECAcomps}
\end{figure*}

The ECA components can be used for structural interpretation \cite{Niskanen2022,Vladyka2023}. We analyze the first component (Figure~\ref{fig:ECAcomps}) in terms of convoluted distributions of interatomic distances, a subset of the entire LMBTR descriptor. This component shows the predominant structural changes associated with spectral ROI variation: increase in ROIs I and III, and decrease in ROI II (see Supplementary Information). There is a striking shift in N1--H, N$_\mathrm{ex}$--C and N$_\mathrm{ex}$--O interatomic distances, and N$_\mathrm{ex}$--N distances, which however remain somewhat inconclusive due to the limited distance range of the descriptor. The surrounding water affects the spectrum as for both N$_\mathrm{ex}$--H and N$_\mathrm{ex}$--O there is a shift at around 3\,{\AA} near the first H$_2$O solvation shell. Interpretation of the angular structural features from the LMBTR is more complicated because exponential distance weighting was found to be required for best emulation performance. The weighting mode was one of the numerous hyperparameters varied in the model selection phase.

\begin{figure*}
	\centering
	\includegraphics[width=\textwidth]{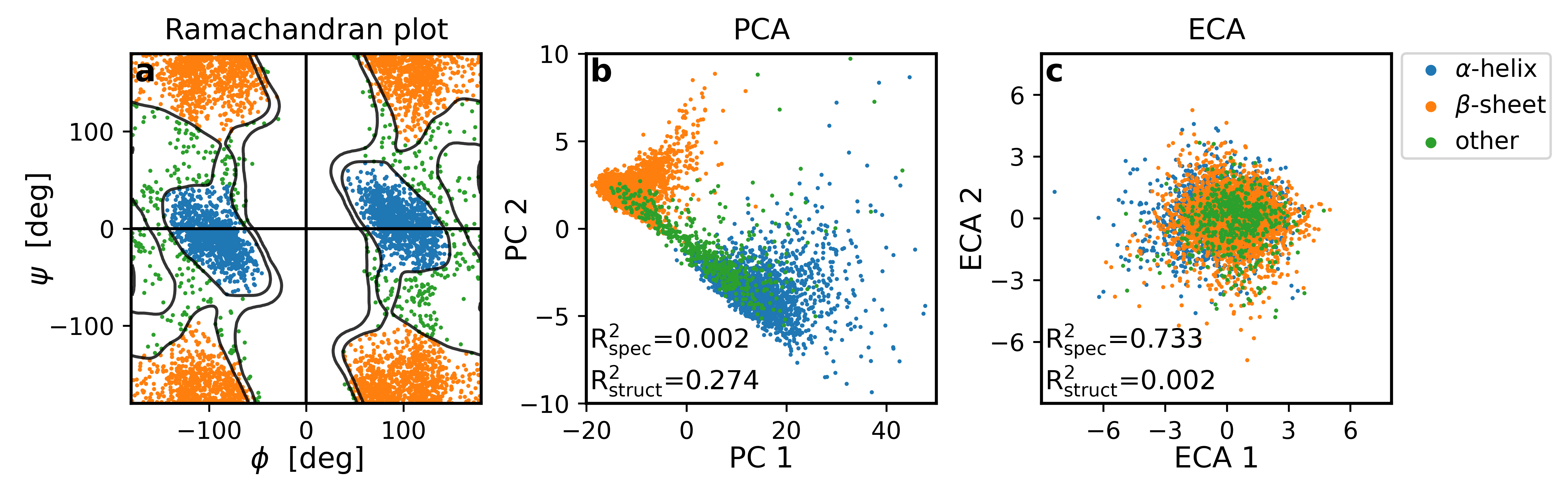}
	\caption{Analysis of the structural classes of the test data with PCA of LMBTR features and ECA reconstruction. {\bf a}: Data points classified to structural classes. The allowed and favored regions according to the Top8000 data set \cite{Hintze2016} are shown as contours. {\bf b}: A 2-component PCA decomposition of the LMBTR features of the data identifies the structural classes. {\bf c}: A 2-component ECA-coordinate reconstruction, based on spectral ROIs, covers drastically more spectral variation but does not distinguish the structural classes. For details, see text.}
	\label{fig:PCAvsECA}
\end{figure*}

We next turn our focus on the features representing the secondary structure, the Ramachandran angles. We assigned each simulated structure in the test set (includes ECA fit and ECA validation data) to one of three classes: $\alpha$-helix, $\beta$-sheet or `other' based on the contours \cite{Lovell2003} from the Top8000 data set \cite{Hintze2016}. Borderline cases were assigned manually as shown in Figure~\ref{fig:PCAvsECA}a. The fractions of $\alpha$-helices and $\beta$-sheets are given in Table \ref{tab:alpha_beta_ratio}. We then used principal component analysis (PCA) to reduce the dimensionality of the LMBTR-encoded structures to two (spectral R$^2 = 0.002$) as shown in Figure~\ref{fig:PCAvsECA}b. This plot shows clear clusters of $\alpha$-helices and $\beta$-sheets with `others' found in between the two. The LMBTR can thus encode this information, even though the dihedrals $\phi$ and $\psi$ are not directly included in it. Correspondingly, we used the ECA approach as a 2D dimensionality reduction tool for the same data (spectral R$^2 = 0.733$), as depicted in Figure~\ref{fig:PCAvsECA}c, without clearly observable clusters. This difference can be explained by PCA focusing only on covered structural variance, whereas ECA is guided by the spectral one. The result therefore indicates that, for reasonable structures, XAS (or XRS) is insensitive to the Ramachandran angles, which in turn cannot be reconstructed from the spectral ROIs.

\begin{table}[h]
	\centering
	\caption{The relative occurence (in \%) of $\alpha$-helices and $\beta$-sheets at 300\,K and 350\,K with the three different force fields.\label{tab:alpha_beta_ratio}}
	\begin{tabular}{p{0.25\linewidth}|cc|cc}
		& \multicolumn{2}{c|}{300 K} &  \multicolumn{2}{c}{350 K}\\
		Force field & $\quad\alpha\quad$ & $\quad\beta\quad$ & $\quad\alpha\quad$ & $\quad\beta\quad$ \\ \hline
		AMBER03   & 31 & 61 & 27 & 62 \\
		Charmm27  & 66 & 30 & 58 & 34\\
		OPLS-AA   & 27 & 64 & 24 & 63 \\
		\hline
		\hline
	\end{tabular}
\end{table}

Feature importance is a metric used for the significance of an input feature with respect to the output of an ML model \cite{Geron2019}. We define an importance score for each LMBTR feature as its absolute value in the first ECA vector. Next, we trained a new emulator with the same architecture as the original one (except for the input layer) using a given number of LMBTR features of the highest importance score. We tested the according models using the ECA validation set and find that most LMBTR groups ({\it e.g.} a pair-wise distance distribution) contain highly meaningful features in terms of performance which, in turn, indicates strong interplay of the structural features of all kinds. Improvement of covered spectral ROI variance is monotonous with respect to the importance score and strongly saturates already with 300 features (see Supplementary Information). This shows that magnitude of a feature in the ECA vector directly measures its spectral significance.

\section{Discussion}
Although the different force fields give different structural distributions (including the those of the Ramachandran angles) the spectral effect predicted upon change of temperature is similar for all of them. This complicates the validation of the force fields by core-level excitation spectra. Moreover, the predicted difference profiles are small and cannot be confirmed or rejected by the current experiment, as all predicted changes are within the error bars that also include the possibility of zero effect. However, the experiment supports the validity of the spectrum simulations, together with the respective convergence checks.

Schwartz and co-workers used a liquid jet to measure the nitrogen K-edge X-ray absorption spectrum of triglycine(aq) in ambient temperature \cite{Schwartz2010} by total electron yield. They obtained a spectrum with roughly similar features, but notably different pre peak to main edge ratio. Interestingly, our study is in agreement with X-ray spectroscopy of solid triglycine \cite{Gordon2003,Zubavichus2004,Weinhardt2019}. The bump at the post edge seen only in our 350\,K spectrum seems to be present in the other studies \cite{Schwartz2010,Gordon2003,Zubavichus2004,Weinhardt2019} and is also present in the N K-edge spectrum of the similar molecule diglycine \cite{Gordon2003}. The cause of this feature is not explained by our simulations or the measured 300\,K spectrum, but based on these references the possibility of some solid triglycine in our 350\,K heated sample cannot be excluded. The aforementioned results have been obtained with numerous yield techniques not necessarily equivalent to XRS used here, or the definition of XAS. Although the XRS has a rather low count rate, it is known to be bulk sensitive and extremely stable.

The descriptor used in an ML work sets the `language' for the resulting scientific discussion. Although several descriptors for the atomistic structure of molecules have been developed \cite{Himanen2020,Huo2022,Behler2011,Bartok2013,Rupp2012,Hansen2015,Chandrasekaran2019}, we see three general requirements for the spectrum analysis by ECA:
\begin{enumerate}
	\item The descriptor must allow for accurate ML with the available data. 
	\item The descriptor must allow for ECA decomposition of high spectral variance coverage with only a few dominant components.
	\item The descriptor must allow for interpretation in terms of structure; preferably it is translatable into simple structural information. 
\end{enumerate}
The architecture of a well-performing descriptor gives insight into the physical system, as it can encode relevant structural information. With LMBTR, the distance-based weighting of angles improved the ML prediction accuracy, which is understandable as the nearby atoms should, by intuition, affect the spectrum more than the distant ones. Unfortunately, the weighting also makes the angular features of the descriptor harder to interpret and (3) is not reached. Thus, effective encoding of physical information for condition (1) may render some of it unrecoverable, and moreover, a trade-off between all three conditions can be expected. Previously, we have applied a similar method for glassy GeO$_2$ \cite{Vladyka2023}, where we found a variant of a Coulomb matrix \cite{Rupp2012} (similar to the Bag of Bonds \cite{Hansen2015}) a suitable descriptor for ML and ECA. Arguably the well-defined covalent bond topology renders LMBTR well-suited for aqueous triglycine.

The ECA algorithm applies projection of the structural descriptor vector onto a limited number of basis vectors and relies on an emulator to predict the spectral information for these new points much faster than the corresponding electron structure calculation would do. This allows for using iterative algorithms for the search of the the optimal basis vectors. The emulator must be complex enough of a function (likely non-linear) yet generalizable to sufficiently capture the relation between any relevant structure $\mathbf{R}$ and its spectrum $\mathbf{S}$. The selection of this mapping $\mathbf{S}_\mathrm{emu}(\mathbf{D}(\mathbf{R}))$ is a complicated problem with plenty of tunable hyperparameters, including those of the descriptor $\mathbf{D}(\mathbf{R})$. In this work we settled for nine independent neural networks, one for each atomic site and ROI with a combined output, as they worked best for the LMBTR-encoded triglycine system with points (1)--(3) in mind. As often the case in machine learning, the choice cannot be shown to be the global optimum, but only the best in the particular hyperparameter search. Up to date there are no generally accepted performance criteria in the X-ray spectroscopic community.

The results of the importance score analysis show that almost all groups of distance and angular distributions are necessary to maximise the covered spectral ROI variance. On the other hand, a significant number of features from each group could be ignored, as only 300 of them in total (originally 1140) are necessary to cover nearly all the accessible variance, which itself rises quickly and monotonously with respect to the number of selected features. This shows that the first ECA vector can indeed find the most relevant features of the descriptor in the order of their spectral significance and could serve as a tool for feature selection. Using the ECA, the dependence of ROI intensities, as captured by the descriptor--emulator--ROI mapping $\mathbf{S}_\mathrm{emu}(\mathbf{D}(\mathbf{R}))$, can practically be condensed into {\it three} structural degrees of freedom when a 1140-dimensional LMBTR is used.

The quantitative analysis of our simulations proves that while information about the Ramachandran angles is present in the LMBTR-encoded structural space, it is largely lost in the subspace covering a significant portion of the spectral ROI variance. Therefore, our results support the conclusion of Schwartz and co-workers \cite{Schwartz2010} in that XAS is insensitive to the secondary structure of triglycine (within the reasonably expected structural space). Furthermore, our results are at least partly in contradiction with that of Gordon and co-workers \cite{Gordon2003} as we see a small dependency between the most relevant part of the backbone conformation and its spectrum (Figure~\ref{fig:PCAvsECA}c) at most. This conclusion is manifested by also taking the mean spectra of all $\alpha$-helices and all $\beta$-sheets in the data (see Supplementary Information), which show a deviating difference profile from those presented in Figure~\ref{fig:ramasimexp}. The advantage of K-edge spectra, locality, seems to be a limitation when it comes to secondary structure of proteins, for which there is an information bottleneck for reasonably expected structures. The ECA shows the ROIs of the total spectrum to mainly be sensitive to the nearest C and N atoms from the solute together with the O and H atoms mostly from the water; and additionally, to the nearest H atoms in the case of N-terminus of the peptide.

\section{Conclusions}
Experimental N K-edge X-ray Raman scattering spectra of aqueous triglycine can be modelled by classical MD and TP-DFT calculations for a good match. However, the experiment is not able to confirm or reject the predicted temperature difference effect. A machine learning emulator can predict the intensities of spectral regions of interest from the corresponding local many-body tensor representation encoded structures. This enabled the application of the emulator-based component analysis, which was able to condense the structure--spectrum dependency practically into three degrees of freedom. Moreover, the spectral significance of structural features was found to be ordered along the magnitude of the according component in the first obtained basis vector. From the analysis we conclude that the details of the secondary structure of aqueous triglycine, expressed by the Ramachandran angles, are lost in X-ray absorption spectral regions of interest due to an information bottleneck. Instead, in the structural information available for analysis from the used descriptor, the distance distributions between the absorption site and its nearest neighbors significantly account for the spectral region variance. Each of the tried force fields results in a similar temperature-difference profile, and therefore distinguishing between them by X-ray absorption seems improbable. On the other hand, the spectrum can be predicted equally well with all of them.

In this work we have pushed structural decomposition by the ECA algorithm to maximal size scales that X-ray spectra can be hoped to have sensitivity to. The method identifies spectrally relevant structural subspace in a complicated system without prior knowledge. Future prospects of this approach include reconstructing the maximum obtainable structural information from the spectra, the first steps of which have already been taken \cite{Vladyka2023}. We also note that the method is not limited to X-ray spectroscopy, but can be applied to a wide variety of inverse problems for which an emulator for the forward problem is known.

\section*{Author contributions}
E.A.E. Data analysis, machine learning, experiment, simulations and writing manuscript. A.V. Data analysis, experiment and writing manuscript. F.G. experiment. C.J.S Experiment, data analysis and writing manuscript. J.N. Research design, experiment, simulations, writing manuscript and funding.

\section*{Conflicts of interest}
There are no conflicts to declare.

\section*{Acknowledgements}
E.A.E. acknowledges Jenny and Antti Wihuri Foundation for funding. E.A.E., A.V. and J.N. acknowledge Academy of Finland for funding via project 331234. The authors acknowledge CSC -- IT Center for Science, Finland, and the FGCI -- Finnish Grid and Cloud Infrastructure for computational resources. We acknowledge the European Synchrotron Radiation Facility (ESRF) for provision of synchrotron radiation facilities. Prof.\,M.\,Mets\"a-Ketel\"a is thanked for sharing his insight in sample environment materials used in biochemical experiments. Dr.\,H.\,M\"uller at the ESRF is thanked for help regarding chemistry lab and sample preparation during the experiment. Dr.\,Y.\,Watier at the ESRF is thanked for support with the sample environment.

\section*{Data availability}
The data and relevant scripts are available in Zenodo: \href{https://zenodo.org/doi/10.5281/zenodo.8239300}{10.5281/zenodo.8239300}.

\bibliographystyle{unsrt}
\bibliography{references}

\newpage
\newpage
\newpage
\clearpage
\onecolumngrid
\section*{Supplementary information}
\subsection*{Consistency of the spectrum simulations}
\begin{figure*}[h!]
    \centering
    \includegraphics[width=\textwidth]{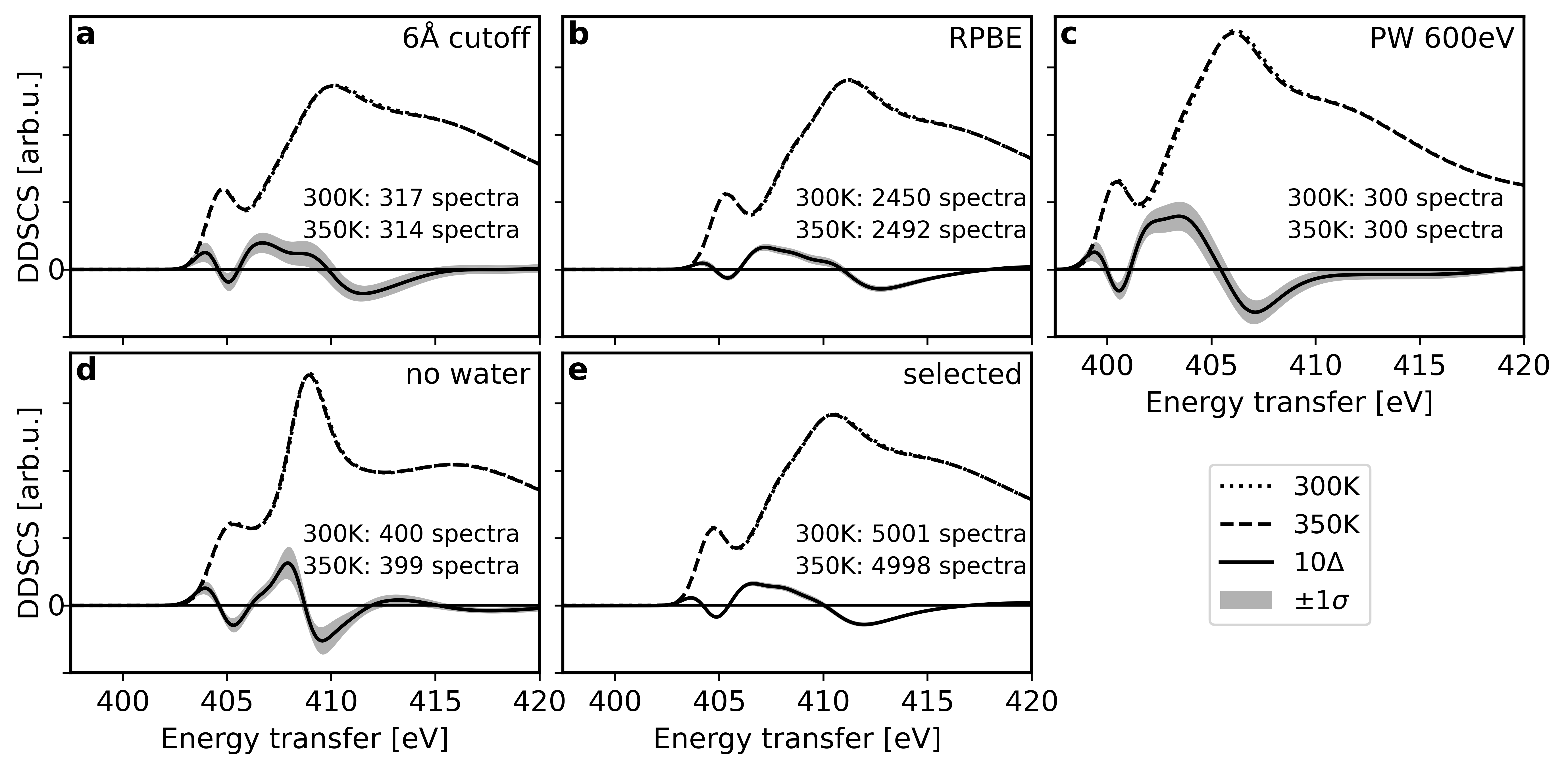}
    \caption{The difference profile remains qualitatively the same with {\bf a}: 6.0\,{\AA} water radial cutoff (one third of the snapshots did not converge), {\bf b}: RPBE functional (4 snapshots did not converge) and {\bf c}: larger plane-wave (PW) energy cutoff (2 snapshots did not converge), apart from an energy shift associated with a more complete basis set. {\bf d}: With the water completely removed the shape of the simulated spectra differ from the experimental and other computational results (1 snapshot did not converge). {\bf e}: The spectra with the selected simulation parameters.}
    \label{fig:consistency}
\end{figure*}

Consistency of the spectrum simulations was studied with respect to geometric water cutoff, the plane-wave energy cutoff, the exchange-correlation potential, and the presence of the solvent. The results are presented in Figure~\ref{fig:consistency}, which indicates similar shape for the obtained difference profile in all cases. The statistical uncertainty $\sigma$ of the difference profile $\Delta$ was calculated using a 10000-fold bootstrap procedure. The calculations suffered from convergence issues: one third of evaluations with 6.0\,{\AA} solvent cutoff did not converge creating a potential selection bias. Furthermore, calculations failed for a few snapshots in the other cases.

\newpage

\subsection*{The best LMBTR hyperparameters within the DScribe package}
A full list of the selected LMBTR hyperparameters within the implementation by DScribe \cite{Himanen2020} is presented in Table \ref{tab:lmbtr}.
\begin{table}[h!!]
\caption{The selected LMBTR parameters within the DScribe package \cite{Himanen2020}.}
\begin{tabular}{c | c | c} 
Parameter & Distances (k2) & Angles (k3)  \\ \hline
geometry function & distance & angle \\
grid min & $0.7~\text{\AA}$ & $0^\circ$\\
grid max & $4.0~\text{\AA}$ & $180^\circ$\\
grid n & 30 & 10\\
grid $\sigma$ & $0.2~\text{\AA}$ & $12.5^\circ$\\
weighting function & unity & exponential\\
weighting scale & --- & 0.8\\
weighting threshold & --- & 1e--7\\
\end{tabular}
\label{tab:lmbtr}
\end{table}
\subsection*{Correlation between structural features and ROI intensities}

\begin{figure*}[hb!]
    \includegraphics[width=0.45\textwidth]{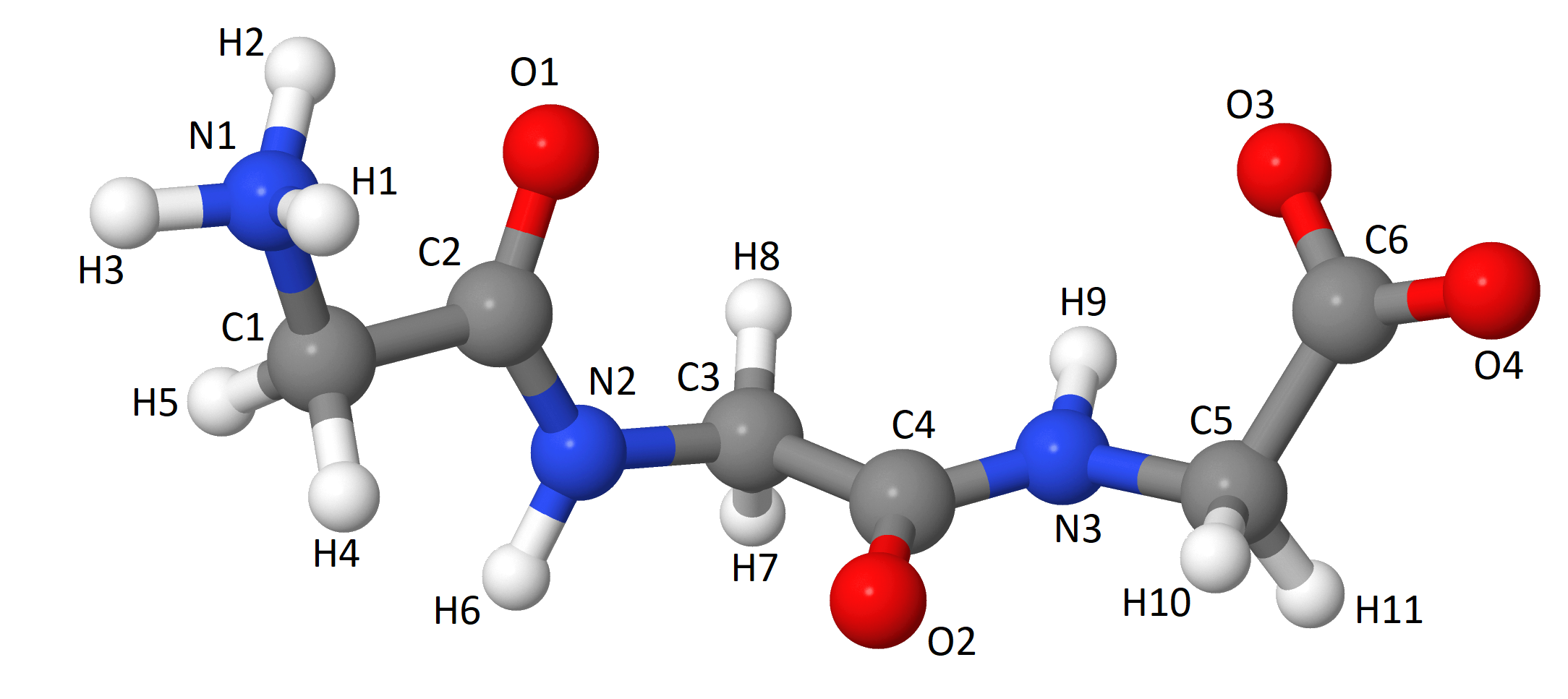}
    \caption{The naming of the atoms. Molecular plot made by the Jmol software.}
    \label{fig:index_meaning}
\end{figure*}

We evaluated Pearson's r coefficients between the ROIs and internal coordinates {\it i.e.} bond lengths $r$, angles $\theta$, and dihedrals $\phi$. In addition, the following water parameters were investigated: number of donated $D$ and accepted $A$ hydrogen bonds, and the number of water molecules in each of the two solvation shell $SS1$ and $SS2$ with respect to the absorbing nitrogen. The definitions of the water parameters are described in \cite{Niskanen2017}. Figure~\ref{fig:index_meaning} shows the naming convention of the atoms in this analysis. 
\par
The results reveal that there are only a few features whose correlation is consistent with respect to the temperature difference profile for all ROIs ({\it i.e.} the sign of the correlation with a ROI intensity is always either equal or opposite to the sign of the difference profile) within 2$\sigma$ error obtained from a 10000-fold bootstrap resampling (Table \ref{tab:concorrs}). There are also many features whose correlation is not strictly inconsistent with respect to the difference profile (Table \ref{tab:corrs}). Overall, the correlation analysis of the internal coordinates and ROIs reveals notable complexity of the problem and necessitates a more sophisticated analysis, that accounts for collaborative action of several structural features.

\begin{table}
\caption{The features whose correlation is consistent with respect to the temperature difference profile with 2$\sigma$ uncertainty. Absolute Pearson's r coefficients greater than or equal to 0.1 are given in bold.}
\begin{tabular}{c | c | c | c } 
 & I & II & III  \\ \hline
$r_\mathrm{C1,C2}$ & $\hphantom{-}0.03$ & $-0.04$ & $\mathbf{\hphantom{-}0.11}$\\
$r_\mathrm{N1,C1}$ & $-0.01$ & $\mathbf{\hphantom{-}0.49}$ & $\mathbf{-0.36}$\\
$r_\mathrm{C4,O2}$ & $\mathbf{\hphantom{-}0.26}$ & $-0.06$ & $\hphantom{-}0.05$\\
$r_\mathrm{C2,O1}$ & $\mathbf{\hphantom{-}0.13}$ & $-0.06$ & $\hphantom{-}0.05$\\
$\theta_\mathrm{C6,O2,O3}$ & $\hphantom{-}0.04$ & $-0.02$ & $\hphantom{-}0.02$\\
$\phi_\mathrm{N3,C5,H11,C6}$ & $-0.02$ & $\hphantom{-}0.03$ & $-0.02$\\
$D_\mathrm{N2}$ & $\hphantom{-}0.04$ & $-0.07$ & $\mathbf{\hphantom{-}0.10}$\\
$D_\mathrm{N3}$ & $\hphantom{-}0.06$ & $-0.07$ & $\hphantom{-}0.09$\\
\end{tabular}
\label{tab:concorrs}
\end{table}

\begin{table}
\caption{The features whose correlation is not inconsistent with respect to the temperature difference profile with 2$\sigma$ uncertainty. Absolute Pearson's r coefficients greater than or equal to 0.1 are given in bold.}
\begin{tabular}{c | c | c | c } 
 & I & II & III  \\ \hline
$r_\mathrm{C6,O4}$ & $\hphantom{-}0.02$ & $\hphantom{-}0.00$ & $-0.01$\\
$r_\mathrm{C4,N3}$ & $\mathbf{\hphantom{-}0.27}$ & $-0.01$ & $\mathbf{\hphantom{-}0.40}$\\
$r_\mathrm{C2,N2}$ & $\mathbf{\hphantom{-}0.10}$ & $\hphantom{-}0.00$ & $\mathbf{\hphantom{-}0.38}$\\
$r_\mathrm{C1,C2}$ & $\hphantom{-}0.03$ & $-0.04$ & $\mathbf{\hphantom{-}0.11}$\\
$r_\mathrm{N1,C1}$ & $-0.01$ & $\mathbf{\hphantom{-}0.49}$ & $\mathbf{-0.36}$\\
$r_\mathrm{O3,C6}$ & $\hphantom{-}0.02$ & $\hphantom{-}0.00$ & $-0.01$\\
$r_\mathrm{C4,O2}$ & $\mathbf{\hphantom{-}0.26}$ & $-0.06$ & $\hphantom{-}0.05$\\
$r_\mathrm{C3,H8}$ & $-0.01$ & $-0.01$ & $\hphantom{-}0.02$\\
$r_\mathrm{C2,O1}$ & $\mathbf{\hphantom{-}0.13}$ & $-0.06$ & $\hphantom{-}0.05$\\
$r_\mathrm{C1,H4}$ & $-0.01$ & $-0.02$ & $\hphantom{-}0.03$\\
$r_\mathrm{C1,H5}$ & $-0.01$ & $-0.01$ & $\hphantom{-}0.02$\\
$r_\mathrm{N1,H1}$ & $-0.00$ & $\mathbf{\hphantom{-}0.17}$ & $-0.03$\\
$r_\mathrm{N1,H2}$ & $\hphantom{-}0.00$ & $\mathbf{\hphantom{-}0.16}$ & $-0.03$\\
$r_\mathrm{N1,H3}$ & $-0.01$ & $\mathbf{\hphantom{-}0.16}$ & $-0.04$\\
$\theta_\mathrm{N2,C3,C4}$ & $\mathbf{-0.19}$ & $\hphantom{-}0.03$ & $\hphantom{-}0.01$\\
$\theta_\mathrm{C1,C2,N2}$ & $-0.03$ & $\hphantom{-}0.01$ & $\hphantom{-}0.00$\\
$\theta_\mathrm{C6,O2,O3}$ & $\hphantom{-}0.04$ & $-0.02$ & $\hphantom{-}0.02$\\
$\theta_\mathrm{C3,H8,C4}$ & $\mathbf{\hphantom{-}0.14}$ & $-0.02$ & $-0.01$\\
$\theta_\mathrm{N2,H6,C3}$ & $-0.01$ & $-0.01$ & $\hphantom{-}0.07$\\
$\theta_\mathrm{N1,H1,C1}$ & $\hphantom{-}0.01$ & $-0.03$ & $\hphantom{-}0.03$\\
$\theta_\mathrm{N1,H2,C1}$ & $-0.00$ & $-0.03$ & $\hphantom{-}0.03$\\
$\theta_\mathrm{N1,H3,C1}$ & $-0.01$ & $-0.03$ & $\hphantom{-}0.02$\\
$\phi_\mathrm{N3,C5,H10,C6}$ & $\hphantom{-}0.01$ & $-0.02$ & $\hphantom{-}0.01$\\
$\phi_\mathrm{N3,C5,H11,C6}$ & $-0.02$ & $\hphantom{-}0.03$ & $-0.02$\\
$\phi_\mathrm{N2,C3,H7,C4}$ & $-0.02$ & $-0.01$ & $-0.00$\\
$\phi_\mathrm{N2,C3,H8,C4}$ & $\hphantom{-}0.01$ & $\hphantom{-}0.01$ & $\hphantom{-}0.00$\\
$\phi_\mathrm{N1,C1,H4,C2}$ & $-0.01$ & $-0.02$ & $\hphantom{-}0.00$\\
$\phi_\mathrm{N1,C1,H5,C2}$ & $\hphantom{-}0.02$ & $\hphantom{-}0.01$ & $\hphantom{-}0.00$\\
$SS1_\mathrm{N1}$ & $-0.01$ & $-0.01$ & $-0.03$\\
$SS1_\mathrm{N2}$ & $\hphantom{-}0.01$ & $\hphantom{-}0.01$ & $-0.02$\\
$SS1_\mathrm{N3}$ & $\hphantom{-}0.06$ & $-0.02$ & $-0.00$\\
$D_\mathrm{N1}$ & $\hphantom{-}0.00$ & $\mathbf{-0.10}$ & $\mathbf{\hphantom{-}0.17}$\\
$D_\mathrm{N2}$ & $\hphantom{-}0.04$ & $-0.07$ & $\mathbf{\hphantom{-}0.10}$\\
$A_\mathrm{N2}$ & $\hphantom{-}0.01$ & $-0.01$ & $\hphantom{-}0.01$\\
$D_\mathrm{N3}$ & $\hphantom{-}0.06$ & $-0.07$ & $\hphantom{-}0.09$\\
$A_\mathrm{N3}$ & $\hphantom{-}0.02$ & $-0.01$ & $\hphantom{-}0.01$\\
\end{tabular}
\label{tab:corrs}
\end{table}
\par

\clearpage
\subsection*{Analysis of the spectral ROI intensity changes along the first ECA component}
\begin{figure}[h!!]
\centering
\includegraphics[width=0.75\textwidth]{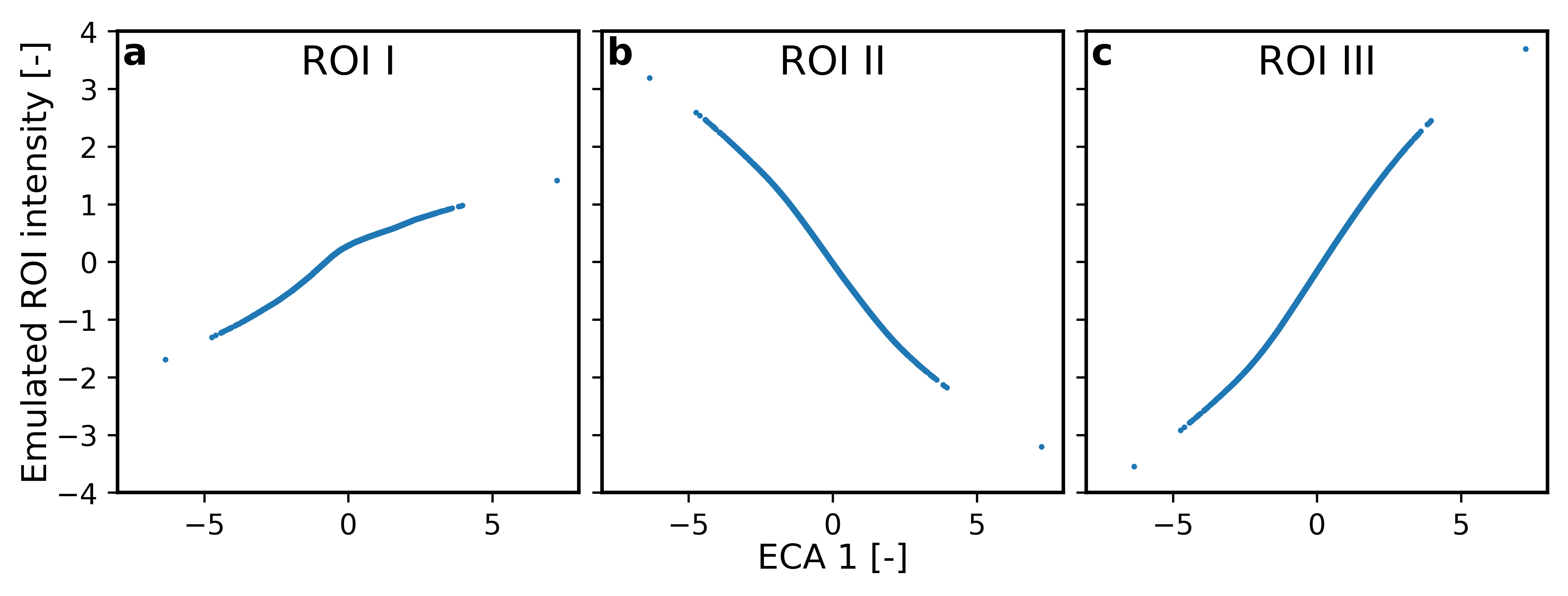}
\caption{Intensity behaviour of the ROIs along the first ECA component {\bf a}--{\bf c}: An increase is observed for ROI I and ROI III, and and a decrease is observed for ROI II.}
\label{FigureS3}
\end{figure}

Figure \ref{FigureS3} shows an increasing trend for the ROI I and III intensities, and a decreasing one for the ROI II intensity, along the first ECA component vector.

\clearpage
\subsection*{Spectral significance of the components in the ECA vector}
We carried out analysis of features by ordering them along the importance score (component magnitude in the first ECA vector). We trained an emulator multiple times iteratively removing 30 of the least important features at each cycle. Only the input layer of the emulator architecture was modified between the iterations.
\par
\begin{figure}[h!!]
\centering
\includegraphics[width=0.95\textwidth]{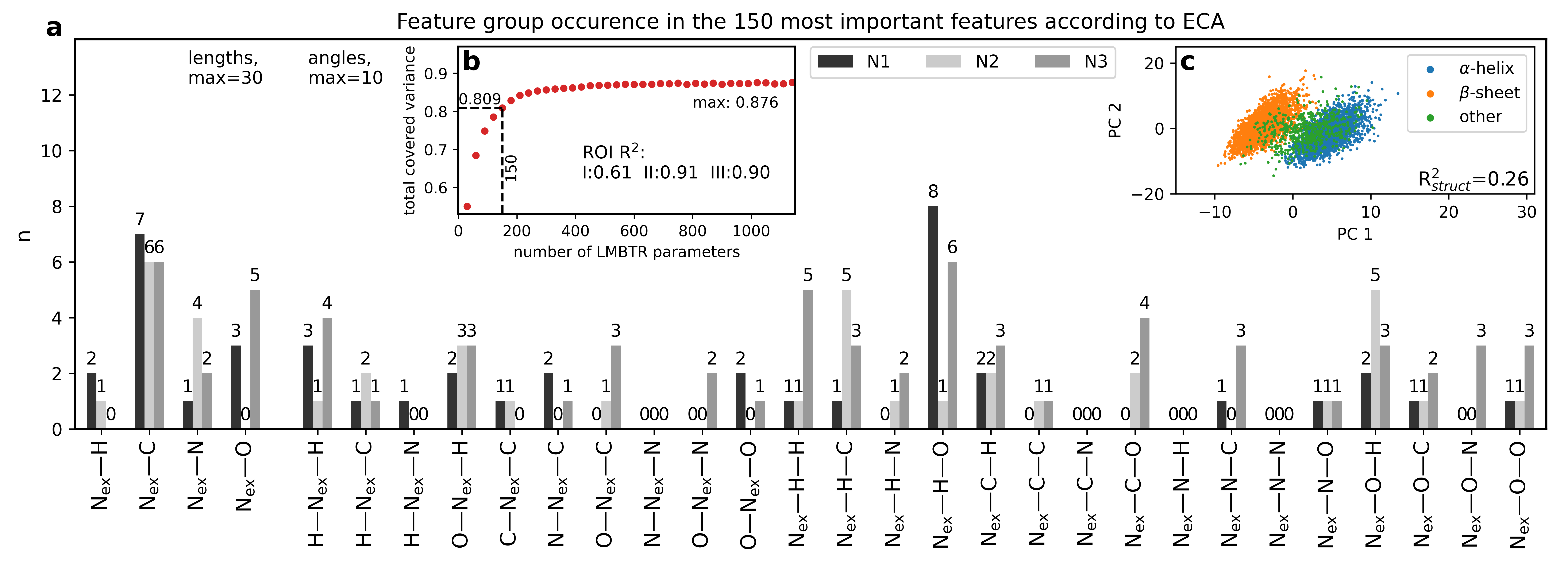}
\caption{{\bf a}: The frequency of occurrence of a feature from different LMBTR groups when 150 features of the highest importance score are used. {\bf b}: The covered spectral ROI intensity variation as a function of the set size. Rapid and monotonous saturation is observed and the set of 150 features covers most of the spectral variance. {\bf c}: 2-component PCA of the 150-dimensional feature vectors from feature selection by the importance score. The information about the Ramachandran angles is still present in this structural descriptor.}
\label{fig:150feats}
\end{figure}
\begin{figure}[h!!]
\centering
\includegraphics[width=0.95\textwidth]{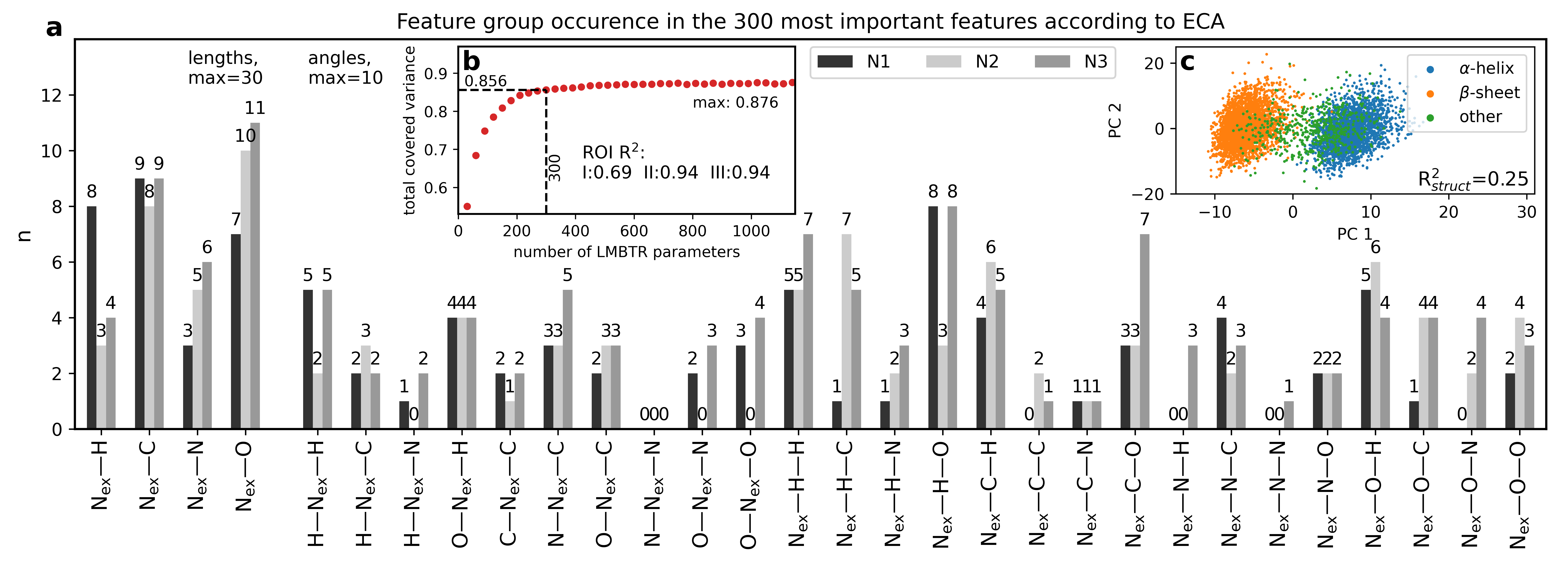}
\caption{{\bf a}: The frequency of occurrence of a feature from different LMBTR groups when 300 features of the highest importance score are used. {\bf b}: The covered spectral ROI intensity variation as a function of the set size. Rapid and monotonous saturation is observed and the set of 300 features covers most of the spectral variance. {\bf c}: 2-component PCA of the 300-dimensional feature vectors from feature selection by the importance score. The information about the Ramachandran angles is still present in this structural descriptor.}
\label{fig:300feats}
\end{figure}

The results show that most and almost all spectral ROI variance is covered by 150 (Figure~\ref{fig:150feats}) and by 300 (Figure~\ref{fig:300feats}) features of the highest importance score, respectively. Almost every group of LMBTR features is present in both cases (Figure~\ref{fig:150feats}a and Figure~\ref{fig:300feats}a). The covered variance rises quickly and monotonously with respect to the number of selected features (Figure~\ref{fig:150feats}b and Figure~\ref{fig:300feats}b), which shows that ECA can find and order the most relevant structural features with respect to the spectral ROI variance. The PCA (Figure~\ref{fig:150feats}c and Figure~\ref{fig:300feats}c) of the selected features shows that the information about the Ramachandran angles is still present in the reduced feature space (150 or 300 features, respectively).

\clearpage
\subsection*{Spectral difference between $\beta$-sheets and $\alpha$-helices}
\begin{figure}[h!!]
    \centering
    \includegraphics[width=0.4\columnwidth]{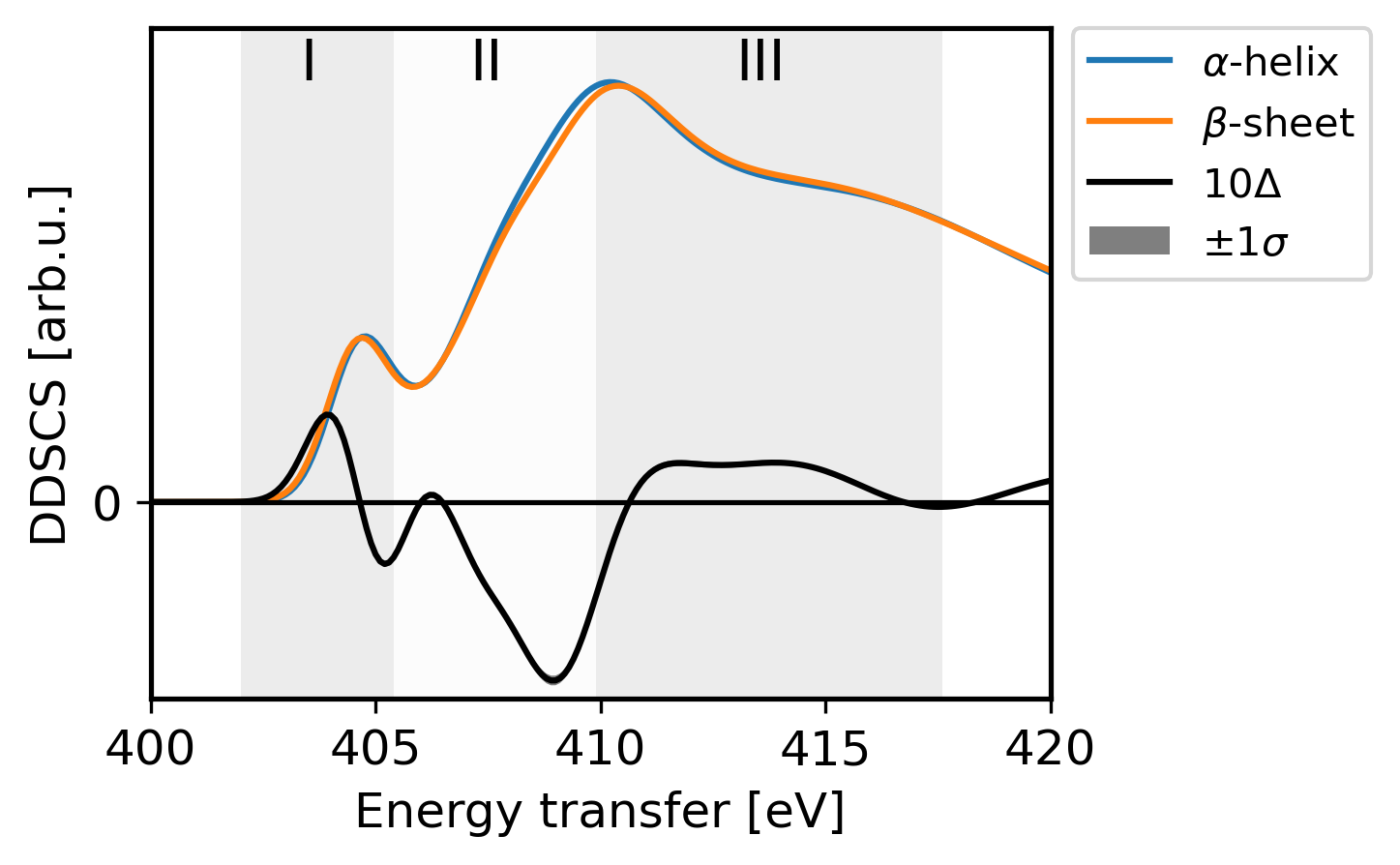}
    \caption{The mean spectra of all the structures classified as $\beta$-sheets or $\alpha$-helices presented together with their difference profile.}
    \label{fig:alpha_vs_beta}
\end{figure}

The mean spectra of all the structures representing $\beta$-sheets or $\alpha$-helices is shown in Figure~\ref{fig:alpha_vs_beta}. The difference profile between the two is also shown. The $\alpha$/$\beta$-difference profile differs drastically from the temperature difference profile.

\subsection*{Evaluating the main results with an alternative convolution}
\begin{figure}[]
    \centering
    \includegraphics[width=0.49\columnwidth]{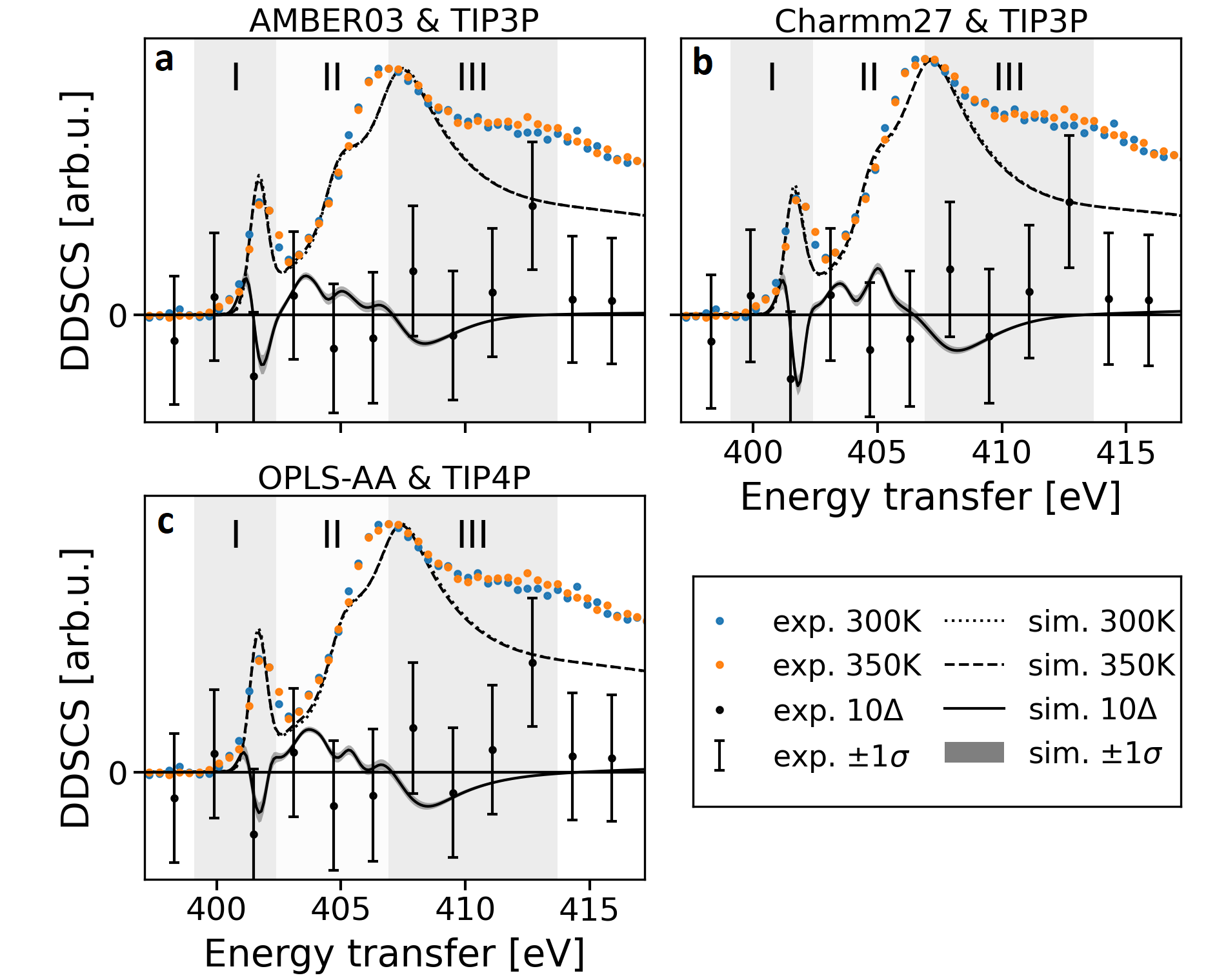}
    \caption{The calculated (and experimental) spectra for aqueous triglycine with alternative convolution parameters. {\bf a}--{\bf c}: Computational ensemble mean and experimental spectra (background removed) with the respective temperature difference profiles $\Delta$ and spectral regions of interest. For the experiment, the difference profile has been 4-fold binned from the spectra. The error bars and shading of the simulated curve indicate the statistical uncertainty $\sigma$ (confidence level 68\%). The computational spectra have been shifted by $-$2.9\,eV in all cases for the pre peak to match with the experiment. The experiment is presented scaled for the same main peak height as the respective 300~K simulation.}
    \label{FigureS7}
\end{figure}

\begin{figure}
    \centering
    \includegraphics[width=\textwidth]{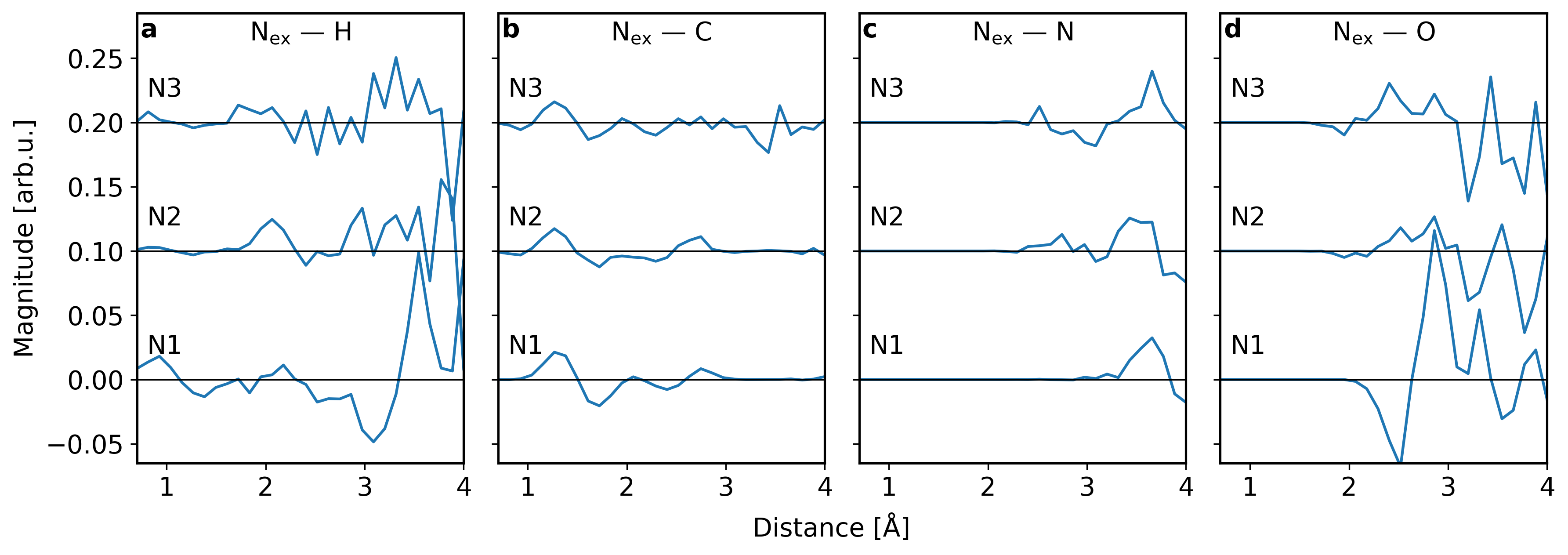}
    \caption{The interatomic distance distributions for each of the nitrogen atoms deduced from the first component vector obtained with an alternative convolution compared to the main text and transformed into the descriptor space. {\bf a}--{\bf d}: Distances from the absorption site (denoted as N$_\mathrm{ex}$) to the neighboring hydrogen, carbon, nitrogen and oxygen atoms, respectively. The ROIs of the total N K-edge spectrum is sensitive to the hydrogen and oxygen atoms at around 3\,{\AA}, which can mainly be contributed to the water molecules. In addition, the ROIs are sensitive to the nearest carbon and nitrogen atoms of each absorption site. Finally, the ROIs are also sensitive to the three nearest hydrogen atoms of N1.
    }
    \label{FigureS8}
\end{figure}

\begin{figure}
    \centering
    \includegraphics[width=\textwidth]{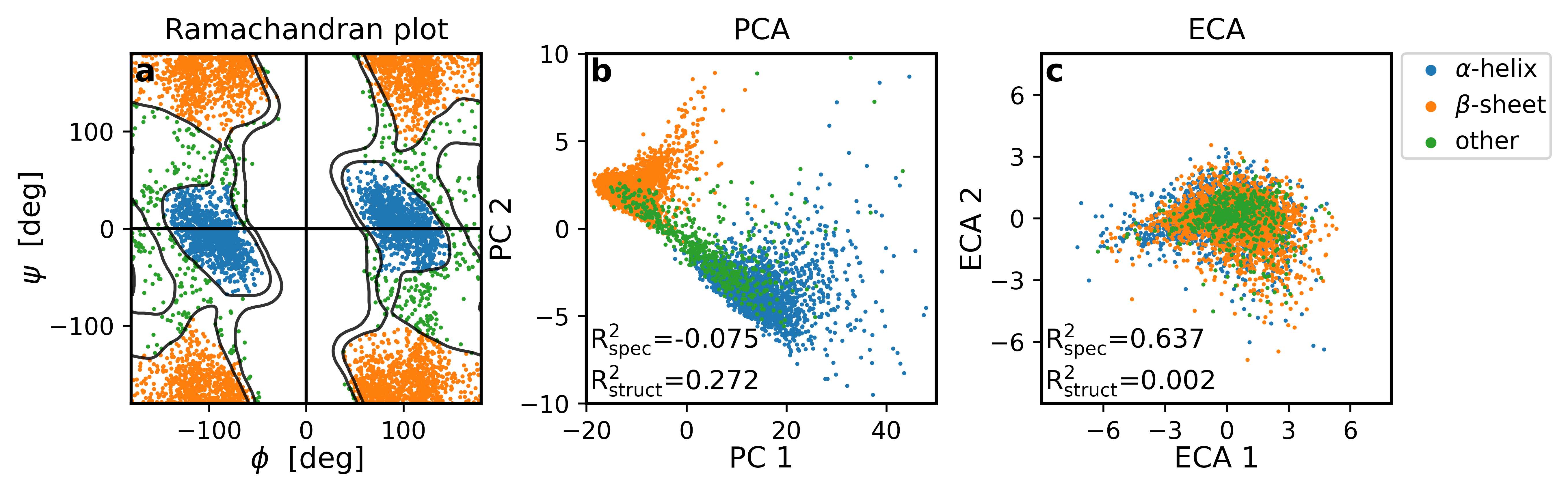}
    \caption{Analysis of the structural classes of the test data with PCA of LMBTR features and ECA reconstruction with an alternative convolution. {\bf a}: Data points classified to structural classes. The allowed and favored regions according to the Top8000 data set \cite{Hintze2016} are shown as contours. {\bf b}: A 2-component PCA decomposition of the LMBTR features of the data identifies the structural classes. {\bf c}: A 2-component ECA-coordinate reconstruction, based on spectral ROIs, covers drastically more spectral variation but does not distinguish the structural classes.}
    \label{FigureS9}
\end{figure}

We evaluated the results with an alternative convolution: Gaussians with FWHM of 0.2\,eV for the lowest state increasing linearly in energy to 4.25\,eV for states 10\,eV above it, or higher. While the shape of the computational spectra differ more drastically from the experimental ones, the difference profile is qualitatively the same (Fig. \ref{FigureS7}). Compared to the main text, the same conclusions can be made from the interatomic distance distributions from the first ECA vector (Fig. \ref{FigureS8}) and the analysis of the $\alpha$/$\beta$ structural classes (Fig. \ref{FigureS9}).

\end{document}